\documentclass[10pt,conference]{IEEEtran}

\usepackage{cite}
\usepackage{amsmath,amssymb,amsfonts}
\usepackage{algorithmic}
\usepackage{graphicx}
\usepackage{textcomp}
\usepackage{xcolor}
\def\BibTeX{{\rm B\kern-.05em{\sc i\kern-.025em b}\kern-.08em
    T\kern-.1667em\lower.7ex\hbox{E}\kern-.125emX}}
\usepackage{booktabs}
\usepackage{multirow}
\usepackage{url}
\usepackage{makecell}
\usepackage{flushend}

\usepackage{xurl} 
\usepackage[breaklinks=true]{hyperref}

\usepackage{pifont}

\newcommand{\greencheck}{\textcolor{green}{\ding{51}}}
\newcommand{\reddash}{\textcolor{red}{\ding{55}}}

\begin{document}
\title{Towards Agentic Honeynet Configuration}

\author{
\IEEEauthorblockN{Federico Mirra, Matteo Boffa, Danilo Giordano, Marco Mellia}
\IEEEauthorblockA{\textit{Politecnico di Torino}\\
\{first.last\}@polito.it}
\and
\IEEEauthorblockN{Idilio Drago}
\IEEEauthorblockA{\textit{Università di Torino}\\
idilio.drago@unito.it}
}
\maketitle

\begin{abstract}
Honeypots are deception systems that emulate vulnerable services to collect threat intelligence. While deploying many honeypots increases the opportunity to observe attacker behaviour, in practise network and computational resources limit the number of honeypots that can be exposed. Hence, practitioners must select the assets to deploy, a decision that is typically made statically despite attacker’ tactics evolving over time.
This work investigates an AI-driven agentic architecture that autonomously manages honeypot exposure in response to ongoing attacks. The proposed agent analyses Intrusion Detection System (IDS) alerts and network state to infer the progression of the attack, identify compromised assets, and predict likely attacker targets. Based on this assessment, the agent dynamically reconfigures the system to maintain attacker engagement while minimizing unnecessary exposure.
The approach is evaluated in a simulated environment where attackers execute Proof-of-Concept exploits for known CVEs. Preliminary results indicate that the agent can effectively infer the intent of the attacker and improve the efficiency of exposure under resource constraints. 
\end{abstract}

\section{Introduction}\label{sec:intro}

Defenders face a persistent asymmetry in cybersecurity: attackers continuously adapt their tactics, discover new vulnerabilities, and exploit weaknesses faster than static defences can evolve. Honeypots -- deceptive systems designed to appear vulnerable and attract attacks -- play a crucial role by collecting threat intelligence and enabling the systematic study of adversarial behaviour~\cite{han2018deception}. However, defenders operate under finite deployment budgets -- limits on computation, bandwidth, or address space -- which restrict the number of honeypots that can be instantiated. Decisions about which services to emulate are typically made statically, resulting in suboptimal alignment with the services attackers actively target. Moreover, static honeypots are easier to fingerprint, as sophisticated adversaries can recognize simple static configurations as decoys. Research on darknet monitoring suggests that dynamic environments sustain attacker interest~\cite{soro2023enlightening}, but operationalizing this strategy requires constant analysis of noisy attacker behaviour at scale, which is impractical to perform manually.

\begin{figure*}[thb]
    \centering
    \includegraphics[
      width=.9\linewidth,
      trim=0.4cm 0.15cm 0.4cm 0.1cm,
      clip
    ]{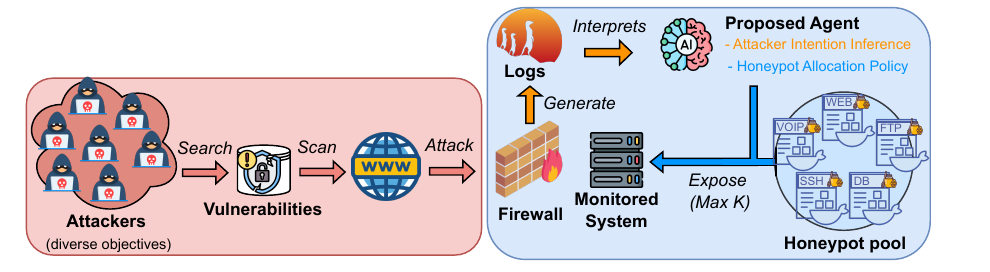}
    \caption{\textbf{Overview of the proposed adaptive honeypot deployment framework.} Attackers actively search for specific vulnerabilities and scan the web to identify machines that satisfy their objectives. The proposed agent interprets attack-related logs, infers the attacker's underlying intentions, and dynamically exposes an optimal subset of honeypots -- subject to a maximum budget of K -- that best match the attacker's goals, enabling effective and adaptive deception.}
    \label{fig:overview_proposal}
\end{figure*}

While recent work has demonstrated that LLMs can generate realistic attacker-facing interactions in honeypots~\cite{sladic2024llm, wang2024honeygpt, otal2024llm, vasilatos2025llmpot}, and that reinforcement learning or game-theoretic methods can optimize honeypot placement~\cite{anwar2022honeypot, huang2019adaptive, guan2023honeyiot}, no existing system uses LLM-based reasoning to autonomously \emph{manage} which services a honeynet exposes over time. In this paper, we present preliminary findings on whether an LLM-based agent can bridge this gap. Figure~\ref{fig:overview_proposal} provides an overview. Attackers possess knowledge of existing vulnerabilities and continuously scan the Internet for systems exposing services aligned with their objectives -- knowledge often unavailable to defenders at deployment time. The key challenge is not merely to observe attacks, but to infer which attack surfaces attackers are actively seeking and deliberately expose them. To this end, we propose an agent that analyses IDS and network logs to extract signals of attacker activity, infers likely attacker goals, and dynamically selects a subset of honeypots from a larger pool. By aligning exposed services with inferred objectives, the system promotes engagement and enables informative attacker interactions under strict resource constraints.

We evaluate the approach using a discrete-state simulation that models adaptive honeypot deployment under adversarial interaction. The simulation instantiates attackers executing proof-of-concept exploits for known CVEs, progressing through a multi-stage exploitation chain aligned with MITRE ATT\&CK: \textit{discovery}, \textit{initial access}, \textit{user-level data exfiltration}, \textit{privilege escalation}, and \textit{root-level data exfiltration}. The defending agent and attackers act in alternating turns, with the agent exposing at most $K$ services per step. An attacker advances only when the agent exposes a service aligned with its current objective. We consider \textit{deterministic}, \textit{probabilistic}, and \textit{consecutive} attacker persistence models, and evaluate the agent across \emph{81 configurations} varying the number of attackers, the pool of vulnerable services, and the LLM backend. Our evaluation focuses on the agent's ability to i) sustain engagement until the attacker completes its chain and ii) accurately track attack progression.

In sum, we make the following contributions. First, we formalize adaptive honeypot exposure under finite budget as an inference-driven decision-making problem. Second, we introduce a discrete-state simulator for evaluating adaptive honeynet management under multi-stage exploitation and diverse attacker models. Third, we propose an autonomous LLM-based agent that reasons over IDS logs to infer attacker intent and selects honeypot services accordingly. Finally, we show that LLM-based agents can track attack progression, infer goals, and adapt deployments to improve intelligence collection under resource constraints. Although obtained in a simplified setting, these results provide preliminary evidence that autonomous honeynet management is feasible.\footnote{Data and source code are available at \url{https://github.com/SmartData-Polito/adaptive-honeynet-agent}}
\section{Problem Statement}\label{sec:problem}

We formally introduce the problem of \emph{adaptive honeypot exposure under finite deployment budget}. A defender seeks to maximize the intelligence gained about attacker behaviour under strict resource constraints. In realistic honeynet deployments, limitations in computation, address space, monitoring capacity, and operational risk prevent all vulnerable services from being exposed simultaneously. Hence, only a limited subset of honeypots can be exposed at any given time. In our deployment, we simulate these constraints by allowing the agent to expose only one honeypot at a time, i.e., only \emph{a single slot} is available to attract attacker interactions.

\hspace{0.1cm} $\bullet$ \textbf{System Model}: Let $\mathcal{H} = \{h_1, h_2, \dots, h_N\}$ denote the set of candidate honeypot services, where each $h_i$ emulates a specific vulnerable application or configuration (e.g., a known CVE). Time progresses in discrete decision epochs $t = 1, 2, \dots$, each corresponding to a control checkpoint at which the defender aggregates recent observations (e.g., IDS alerts and network logs) and may update the exposed services. At each epoch, the defender can expose at most $K \ll N$ services, reflecting a fixed deployment budget.

Attackers interact with the honeynet by scanning for exposed services and attempting exploits aligned with their objectives. We assume attacker behaviour follows a structured, multi-stage intrusion process consistent with the MITRE ATT\&CK framework, progressing through ordered phases such as reconnaissance, initial access, post-exploitation, privilege escalation, and data exfiltration. The progression to later stages depends on the availability of services aligned with the current objective of the attacker.

\hspace{0.1cm} $\bullet$ \textbf{Partial Observability and Intent Inference}: The internal state of the attacker, including goals, progress along the intrusion chain, and future plans, is not directly observable. Instead, the defender observes partial and noisy signals derived from network telemetry and intrusion detection systems (IDS), such as alerts, signatures, and protocol-level indicators. These observations may be incomplete, ambiguous, delayed, or affected by false positives.

Consequently, the defender must infer the attacker’s latent state from the accumulated evidence, mapping low-level indicators to a higher-level understanding of the progression and intent of the attack. This inference problem is inherently temporal as the interpretation of new observations depends on previously observed activity.

\hspace{0.1cm} $\bullet$ \textbf{Aggregation of Concurrent Attack Activity}: In practice, honeynets are subjected to high-volume continuous traffic from multiple independent sources. Scanning, probing, and exploitation attempts from different attackers may overlap in time and interleave between services, generating heterogeneous IDS alerts. Precisely attributing each event to a different attacker or campaign is often infeasible. In this work, we assume that raw network and IDS data are \emph{pre-aggregated} over each decision epoch, and that the agent receives a consolidated stream of attack-related observations reflecting concurrent malicious activity. The agent does not attempt to disentangle individual attackers. Instead, it reasons over a \emph{single aggregated attack process}, inferring a dominant or representative attack progression that reflects the objectives currently exercised against the honeynet. This abstraction aggregates multiple concurrent attackers into a single semantic ``thread of attack'' capturing common exploitation patterns, targeted services, and stages of intrusion.

In our experimental evaluation, this aggregated attack process is instantiated by simulating a single attacker at a time, which serves as an equivalent representative of the concurrent attack flow described above. This modelling choice aligns with the defender’s objective in a honeypot setting: not precise attacker attribution, but identification of the attack surfaces that are actively targeted and their current stages of exploitation. From the perspective of adaptive honeypot exposure, multiple attackers pursuing similar vulnerabilities exert equivalent pressure on the deployment and motivate the same exposure decisions. 

\hspace{0.1cm} $\bullet$ \textbf{Decision Problem}: At each decision epoch $t$, the defender selects a subset $\mathcal{E}_t \subseteq \mathcal{H}$ with $|\mathcal{E}_t| \leq K$ to expose. Exposing services aligned with attacker objectives enables continued progression and reveals additional behaviour, while exposing irrelevant services may delay the attack or cause disengagement. The defender’s objective is not to prevent compromise, but to \emph{maximize intelligence collection} by sustaining engagement under minimal and targeted exposure, inducing a trade-off between exploration and exploitation.

\hspace{0.1cm} $\bullet$ \textbf{Problem Formulation}: We formalize adaptive honeypot exposure as an \emph{inference-driven sequential decision-making problem under partial observability}. At each decision epoch, the defender observes an accumulated stream of network and intrusion-detection events, reflecting both repeated interactions by the same attacker and overlapping activity from multiple attackers. These observations are inherently noisy and indirect, as they do not explicitly encode the current stage or intent of the attacker. Given this history, the defender must:

\begin{enumerate}
\item Infer the latent stage of the ongoing attack process and the likely adversarial objectives from the aggregated partial evidence.
\item Select a budget-constrained subset of services to expose in order to maximize expected future information gain.
\end{enumerate}

This problem is characterized by three key challenges: (i) \emph{intent inference} from low-level and temporally aggregated security telemetry, (ii) \emph{budget-constrained adaptation} of the exposed attack surface under uncertainty, and (iii) \emph{temporal consistency}, as premature hiding or delayed exposure of services can disrupt attack progression.

Manually configured honeypots cannot address these challenges, motivating autonomous agents capable of maintaining beliefs about attack progression, reasoning over accumulated evidence, and dynamically reshaping the honeynet. 
\section{Adaptive Honeynet Management Agent}\label{sec:agent}
We propose an autonomous agent for adaptive honeypot deployment that continuously observes attacker activity, infers adversarial intent, and dynamically adjusts the exposed attack surface under a finite deployment budget. The agent follows a perception--inference--action loop, with a large language model (LLM) acting as the central reasoning component. By abstracting low-level network telemetry into semantically meaningful signals and enforcing decisions through programmable network controls, the system enables closed-loop adaptation without human intervention.

\subsection{Design Principles}

The architecture is guided by three key principles.

\vspace{0.1cm}$\bullet$ \textbf{Evidence-driven reasoning}: All inferences are grounded in observable network evidence. Instead of processing raw packet traces, which would saturate the LLM's context, the agent consumes structured alerts produced by an intrusion detection system (IDS). This ensures scalability while preserving the semantics of the attack.

\vspace{0.1cm}$\bullet$ \textbf{Temporal consistency}: Attacks are modelled as multi-stage processes aligned with the MITRE ATT\&CK framework. The agent reasons over \emph{accumulated, partial, and noisy} observations -- gathered from repeated interactions and overlapping attack activity -- to infer the attacker’s latent current stage and anticipate likely subsequent objectives, maintaining coherent exposure decisions over time.

\vspace{0.1cm}$\bullet$ \textbf{Controlled actuation}: Adaptation is constrained by an explicit exposure budget \(K\), which limits the number of simultaneously exposed honeypots. Exposure decisions are enforced through deterministic network reconfiguration (e.g., firewall rules), ensuring precise, auditable, and reversible actions.

\subsection{Agent Architecture}

The agent operates as a recurring control loop of four logical components (orange and blue boxes of Figure~\ref{fig:simulation_overview}). 

\vspace{0.1cm}$\bullet$ A \textbf{perception layer} periodically collects IDS alerts, honeypot metadata, and the current state of network exposure. IDS alerts compact high-volume traffic into discrete events such as scans, exploit attempts, or post-exploitation activity, making them suitable for semantic analysis.

\vspace{0.1cm}$\bullet$ An \textbf{attack inference layer} uses the LLM to interpret newly observed alerts in the context of previous observations. This step infers the progression of the attacker through a multi-stage exploitation chain, producing an evolving internal representation that captures the attack phases, the targeted services, and the estimated depth of exploitation.

\vspace{0.1cm}$\bullet$ An \textbf{exposure planning layer} also uses the LLM to select a subset of honeypots to expose during the next decision epoch, subject to budget constraints \(K\). The planner prioritizes services aligned with the inferred attacker objective while minimizing unnecessary exposure of unrelated assets.

\vspace{0.1cm}$\bullet$ An \textbf{enforcement and memory layer} translates exposure decisions into concrete network actions (e.g., firewall updates) and records observations, inferences, and actions in episodic memory. This memory provides a short-term context for subsequent reasoning and supports post-hoc analysis.

\subsection{LLM-Centered Reasoning}

The LLM serves as the central reasoning component of the agent: it interprets observed attacker activity and selects appropriate exposure actions. Rather than applying a simple rule-based approach, the model grounds its decisions on contextual information accumulated across multiple iterations. At each decision step, the LLM is provided with (i) summarised IDS alerts from the current epoch, (ii) the previously inferred attack progression, and (iii) the deployment constraints imposed by the exposure budget. The inferred attack progression is represented as an \emph{attack graph}, where nodes correspond to attack stages or compromised services, and edges represent plausible transitions induced by attacker actions. Each attacker interaction contributes evidence for one or more edges in this graph, potentially extending existing branches or activating new ones. The agent does not observe this graph directly; instead, the LLM incrementally infers and updates its structure by reasoning over accumulated alerts, effectively reconstructing which objectives are being pursued and along which exploitation paths. Based on the updated attack graph, the LLM refines the inferred attack state and determines which honeypot services should be exposed next.

Notice that, as the reasoning logic is encapsulated in the LLM, the same architecture can be instantiated with a different model backend, enabling direct evaluation of how model capacity affects the agent's capabilities.

To summarise, the key architectural insight we propose is the separation between \emph{semantic reasoning} -- comprising attack inference and exposure planning, handled by the LLM -- and \emph{deterministic control} -- comprising perception, enforcement, and memory, implemented via conventional networking mechanisms. This separation enables autonomous, adaptive honeynet management while preserving safety, traceability, and strict control over resource usage.
\section{Simulation Environment}\label{sec:simulation}
To evaluate the proposed adaptive honeypot management agent in a controlled, yet adversarial setting, we design a discrete-time simulation that models repeated interactions between attackers and a resource-constrained defender. The simulation captures the temporal dynamics of multi-stage intrusions and allows systematic evaluation of inference accuracy and exposure efficiency under varying attacker behaviours.

\begin{figure}[!t]
    \centering
    \includegraphics[
      width=.9\linewidth,
      trim=0.5cm 0.2cm 0.2cm 0.2cm,
      clip
    ]{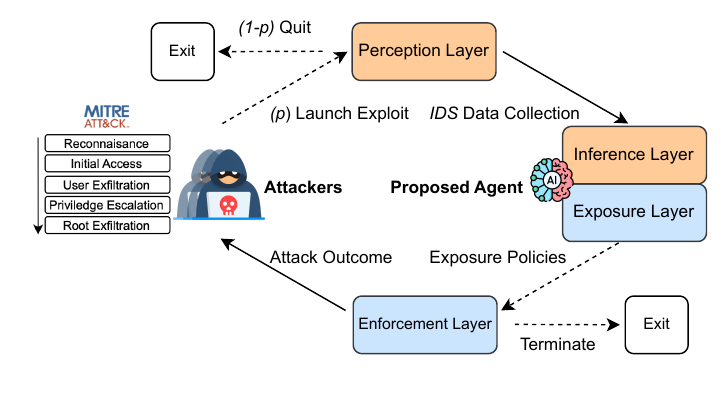}
    \caption{\textbf{Discrete-time simulation of attacker--defender interaction.} The system evolves in epochs alternating between attacker actions and agent decisions, where the agent infers attack progression from IDS data and applies budget-constrained exposure policies.}
    \label{fig:simulation_overview}
\end{figure}

\hspace{0.1cm} $\bullet$ \textbf{Discrete-Time Interaction Model}: The simulation proceeds in discrete epochs, each consisting of an attacker phase followed by a defender decision phase, as illustrated in Figure~\ref{fig:simulation_overview}. At every epoch, the defending agent may expose at most \(K\) honeypot services selected from a larger pool of candidates. By restricting the number of concurrently exposed honeypots (only one exposed honeypot in our scenario), the simulation models realistic operational constraints that forces the agent to dynamically adapt its deception strategy rather than relying on a static, fully exposed attack surface.

During the attacker phase, an adversary scans the currently exposed services and decides whether to interact with them based on their objectives and persistence strategy. In the defender phase, the agent analyses the resulting network and IDS logs, updates its internal representation of the attack progression, and selects the honeypots to expose in the next epoch. We simulate a queue of $A$ attackers, that subsequently interacts in the simulation.

\hspace{0.1cm} $\bullet$ \textbf{Attack Progression Model}: We model attacker behaviour as a multi-stage exploitation chain aligned with the MITRE ATT\&CK framework. Attackers progress through an ordered sequence of phases -- reconnaissance (service scanning), initial access, user-level data exfiltration, privilege escalation, and root-level data exfiltration -- but may terminate the attack upon achieving an intermediate objective (e.g., stopping after user-level exfiltration rather than pursuing full root compromise). The progression to a subsequent phase occurs only when the defender exposes a honeypot service aligned with the attacker’s current objective, directly coupling attacker evolution with exposure decisions. This formulation enables evaluation of whether the agent can infer attacker intent from observations and adapt the exposed attack surface to sustain engagement. The simulator records the progression of the attacker in the ground-truth, which is used to assess the accuracy of the agent’s inferred attack state.

\hspace{0.1cm} $\bullet$ \textbf{Attacker Persistence Models}: To capture various adversarial behaviours, the simulator instantiates multiple persistence strategies of attackers. We consider: (i) \emph{deterministic attackers}, which always attempt exploitation whenever a relevant service is exposed ($p=1$); (ii) \emph{consecutive attackers}, which require the same service to remain exposed across consecutive epochs to progress ($p=1$ if the correct service is exposed, $p=0$ otherwise); and (iii) \emph{probabilistic attackers}, whose likelihood of continuing an attack decreases when exposure is intermittent. More formally, if $g$ is the number of non-exposure epochs, we compute $p$ as:
\[
    \mathrm{p} \;=\;
    \begin{cases}
    1, & \text{if } g\leq0\\[2pt]
    max(p_{min}, 1-d\cdot g), & \text{if } g>0
    \end{cases}
\]
Where $d$ and $p_{min}$ are hyper-parameters representing a decay factor per-epoch and the minimum attempt probability floor.

\hspace{0.1cm} $\bullet$ \textbf{Observations and Termination}: All attacker actions generate synthetic network traces and IDS alerts, which constitute the sole observations available to the defending agent. The agent does not have direct access to the attacker’s internal state or the progression of the ground-truth. A simulation run ends when the agent has processed all attackers. This means, for each attacker, either i) by determining that the inferred attack chain is exhausted, ii) by the attacker abandoning the interaction, or iii) when reaching a predefined epoch limit.

\section{Evaluation Results}\label{sec:results}
We now describe the experimental setup and present the results of the agents for the proposed simulations.

\subsection{Experimental Settings}

\hspace{0.1cm}\textbf{Honeynet configurations}: We consider three simulated scenarios that progressively increase both the ambiguity of the attack-surface and the diversity of attacker objectives.
Let $A$ denote the pool of attackers, $H$ the pool of available machines, and $K$ the number of vulnerable machines that can be deployed per epoch.
Throughout the experiments, we fix $K=1$ and define the following configurations:
\begin{enumerate}
    \item \textbf{Fully vulnerable deployment.} $A=4$ attackers and $H=4$ machines, all of which are vulnerable. This represents the simplest setting, as any machine exposed by the agent can satisfy the objective of an attacker.
    \item \textbf{Small mixed deployment.} $A=2$ attackers and $H=4$ machines, comprising two vulnerable machines and two non-exploitable ones. This setting is more challenging, as the agent must infer that some exposed machines are irrelevant to the attacker.
    \item \textbf{Large mixed deployment.} $A=2$ attackers operate over $H=6$ machines, including two vulnerable machines and four non-exploitable ones. This is the most challenging scenario, as only a small subset of available honeypots is relevant to the attacker’s objectives.
\end{enumerate}
As $|H|$ increases while the number of vulnerable machines remains fixed, the agent must infer the intent of the attacker from an increasingly large and noisy set of candidates.

\begin{figure}[!t]
    \centering
    \includegraphics[
      width=1\linewidth,
      trim=0.75cm 0.2cm 0.75cm 0.2cm,
      clip
    ]{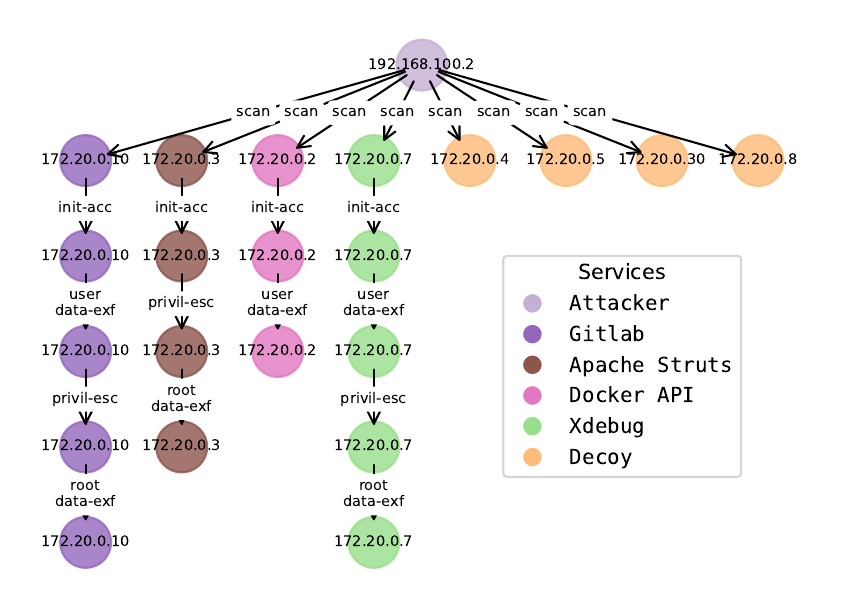}
    \caption{\textbf{Ground-truth attack graph used in the simulation.} Nodes represent attack stages and services, and edges encode feasible progression paths aligned with the MITRE ATT\&CK framework.}
    \label{fig:attack_graph}
\end{figure}

\hspace{0.1cm}\textbf{Attackers and language models}: To simulate realistic adversarial behavior, attackers execute scripted multi-stage exploits derived from public proof-of-concept attacks. These specific attack chains target vulnerabilities in GitLab, Apache Struts, Docker API, and Xdebug, progressing through distinct phases from Initial Access to Root Data Exfiltration. We represent the ground-truth attack graph in Figure~\ref{fig:attack_graph}.

As described in Section~\ref{sec:simulation}, we evaluate three persistence strategies of attackers.
For probabilistic attackers, we set $d=0.25$ and $p_{\min}=0.1$.

To assess the impact of reasoning capacity on inference and exposure control, we evaluate three language-model backends: \texttt{GPT-4.1}, \texttt{GPT-4.1-mini}, and \texttt{gpt-oss-120b}. The first two are accessed via the OpenAI API, whereas \texttt{gpt-oss-120b} is deployed locally. While fine-tuning \texttt{gpt-oss-120b} could potentially improve performance while retaining the advantages of a smaller self-hosted model, this lies outside the scope of the present work; therefore, all models are evaluated in their off-the-shelf configuration.

In the sake of space, we provide a more detailed description of the vulnerable environments, prompt templates, model configurations, and examples of generated IDS alerts in our GitHub repository.

\hspace{0.1cm}\textbf{Evaluation protocol}: Simulations proceed in discrete time with a maximum horizon of 20 epochs.
We evaluate the agent's capability to gather intelligence along two complementary dimensions:
\begin{itemize}
\item \textbf{Exploitation Achieved}: A binary metric that indicates whether agent exposure decisions allow the attacker to reach the terminal node of the ground-truth attack graph.
\item \textbf{Attack-Stage Inference Score}: At each epoch, the agent predicts the current stage of the attacker in the attack graph. During an episode, we compute true positives (TP), false negatives (FN), and false positives (FP) based on agreement with the ground-truth progression and compute $score=\frac{TP}{TP+FP+FN}$.
\end{itemize}

To ensure comparability across scenarios, metrics are computed for the two attackers present in all configurations (\textit{GitLab} and \textit{Apache Struts} of Figure~\ref{fig:attack_graph}).
Each configuration is repeated over three random seeds, resulting in 81 runs in total (3 language models $\times$ 3 honeynet configurations $\times$ 3 persistence strategies $\times$ 3 seeds).

\subsection{Results}

\hspace{0.1cm}\textbf{Exploitation vs. deployment setting}: Table~\ref{tab:exploitation_results} reports exploitation achieved across models and honeynet configurations, with each entry aggregating 9 runs.
In the \emph{fully vulnerable} setting, all models achieve a 100\% success rate, confirming that in the absence of attack-surface ambiguity, the agent reliably sustains attacker progression.

As non-exploitable decoys are introduced, exploitation becomes progressively more difficult.
In the \emph{small mixed} configuration, success rates decrease across all models, indicating that even limited ambiguity challenges exposure control.
This effect is amplified in the \emph{large mixed} configuration, where the proportion of relevant machines is lowest and the exploitation success drops substantially, particularly for \texttt{GPT-4.1-mini}.
In general, these results highlight the central role of attack-surface ambiguity in hindering sustained attacker progression.

\begin{table}[ht]
\centering
\renewcommand{\arraystretch}{1.2}
\caption{Exploitation success rates -- LLM vs deployment settings.}
\label{tab:exploitation_results}
\begin{tabular}{llc}
\hline
\textbf{Model} & \textbf{Deployment setting} & \textbf{Exploitation achieved} \\
\hline
\multirow{3}{*}{GPT-4.1}
 & Fully vulnerable & 9/9 (100\%) \\
 & Small mixed      & 7/9 (78\%) \\
 & Large mixed      & 6/9 (67\%) \\
\hline
\multirow{3}{*}{GPT-4.1-mini}
 & Fully vulnerable & 9/9 (100\%) \\
 & Small mixed      & 8/9 (89\%) \\
 & Large mixed      & 4/9 (44\%) \\
\hline
\multirow{3}{*}{gpt-oss-120b}
 & Fully vulnerable & 9/9 (100\%) \\
 & Small mixed      & 8/9 (89\%) \\
 & Large mixed      & 8/9 (89\%) \\
\hline
\end{tabular}
\end{table}

Notably, \texttt{gpt-oss-120b} appears to outperform larger models in mixed deployments.
As discussed in the following, this apparent advantage is misleading, as the success of the exploitation alone does not reflect the accuracy of the agent’s internal estimate of the attacker's progression.

\hspace{0.1cm}\textbf{Exploitation vs. attacker persistence}: Table~\ref{tab:exploitation_persistence} reports the exploitation achieved as a function of attacker persistence, aggregating the results over deployment settings.
Deterministic attackers are the easiest to maintain and yield consistently high exploitation rates across models.
Probabilistic attackers introduce additional uncertainty, resulting in moderate performance degradation.
The most challenging setting is the \emph{consecutive} persistence model, which requires uninterrupted correct exposure decisions.
In this regime, the success of the exploitation decreases markedly for \texttt{GPT-4.1} and \texttt{GPT-4.1-mini}, indicating the sensitivity to compounding inference errors.
In contrast, \texttt{gpt-oss-120b} achieves perfect exploitation in this setting.
However, a closer inspection reveals that this performance arises from a systematic bias in how the model interprets attacker progression.
While the agent prompt requires mapping IDS evidence to a sequence of attack stages -- from reconnaissance to root-level data exfiltration -- \texttt{gpt-oss-120b} frequently predicts early reconnaissance and final exfiltration stages while largely ignoring intermediate phases.
As a result, the agent exposes machines as if the attacker were near completion, even when the attacker is still progressing through earlier stages, inadvertently facilitating exploitation despite inaccurate internal state estimation.

\begin{table}[ht]
\centering
\renewcommand{\arraystretch}{1.2}
\caption{Exploitation success rates -- LLM vs persistence models.}
\label{tab:exploitation_persistence}
\begin{tabular}{llc}
\hline
\textbf{Model} & \textbf{Attacker mode} & \textbf{Exploitation achieved} \\
\hline
\multirow{3}{*}{GPT-4.1}
 & Deterministic & 9/9 (100\%) \\
 & Probabilistic & 7/9 (78\%) \\
 & Consecutive   & 6/9 (67\%) \\
\hline
\multirow{3}{*}{GPT-4.1-mini}
 & Deterministic & 8/9 (89\%) \\
 & Probabilistic & 8/9 (89\%) \\
 & Consecutive   & 5/9 (56\%) \\
\hline
\multirow{3}{*}{gpt-oss-120b}
 & Deterministic & 8/9 (89\%) \\
 & Probabilistic & 8/9 (89\%) \\
 & Consecutive   & 9/9 (100\%) \\
\hline
\end{tabular}
\end{table}

\hspace{0.1cm}\textbf{Attack inference accuracy}: Finally, Table~\ref{tab:attack_inference_accuracy_full} reports the attack-stage inference score in models, deployment settings, and persistence strategies.
In contrast to exploitation success, the inference score degrades consistently as the ambiguity of the attack-surface increases.
All models perform best in the fully vulnerable setting, with the score decreasing in the small mixed configuration and further dropping in the large mixed scenario.
Across all conditions, \texttt{GPT-4.1} achieves the highest inference score, followed by \texttt{GPT-4.1-mini}, while \texttt{gpt-oss-120b} consistently lags behind.
This holds even in scenarios where \texttt{gpt-oss-120b} exhibits strong exploitation performance.
By prematurely hallucinating evidence from the terminal-stage without correctly tracking intermediate attack phases, \texttt{gpt-oss-120b} can sustain attacker progression while failing to accurately estimate the attacker’s true position in the ground-truth attack graph.
In contrast, \texttt{GPT-4.1} and \texttt{GPT-4.1-mini} exhibit more fine-grained phase tracking, improving the score at the cost of increased sensitivity to exposure errors under high ambiguity.
These results demonstrate that exploitation success alone can be misleading and underscore the importance of jointly evaluating deception effectiveness and attack-stage inference fidelity.

\begin{table}[ht]
\centering
\scriptsize
\renewcommand{\arraystretch}{1.10}
\caption{Attack-Stage Inference Score across LLMs and deployment settings. We aggregated results over random seeds.}
\label{tab:attack_inference_accuracy_full}
\begin{tabular}{llccc}
\toprule
\textbf{Deployment} & \textbf{Attack mode} 
 & \textbf{GPT-4.1} 
 & \textbf{GPT-4.1-mini} 
 & \textbf{gpt-oss-120b} \\
\midrule
\multirow{3}{*}{Fully vuln.}
 & Deterministic & 85.9 $\pm$ 4.2 & 85.2 $\pm$ 10.5 & 55.6 $\pm$ 0.0 \\
 & Probabilistic & 88.9 $\pm$ 0.0 & 77.8 $\pm$ 0.0  & 59.3 $\pm$ 5.2 \\
 & Consecutive   & 88.9 $\pm$ 0.0 & 70.4 $\pm$ 5.2  & 51.9 $\pm$ 5.2 \\
\midrule
\multirow{3}{*}{Small mix.}
 & Deterministic & 83.0 $\pm$ 4.2 & 77.8 $\pm$ 0.0  & 59.3 $\pm$ 5.2 \\
 & Probabilistic & 81.5 $\pm$ 10.5 & 74.1 $\pm$ 5.2  & 50.0 $\pm$ 4.5 \\
 & Consecutive   & 81.5 $\pm$ 10.5 & 74.1 $\pm$ 13.9  & 55.6 $\pm$ 0.0 \\
\midrule
\multirow{3}{*}{Large mix.}
 & Deterministic & 83.0 $\pm$ 4.2 & 77.8 $\pm$ 9.1  & 55.6 $\pm$ 0.0 \\
 & Probabilistic & 81.5 $\pm$ 10.5 & 74.1 $\pm$ 13.9  & 55.6 $\pm$ 0.0 \\
 & Consecutive   & 74.1 $\pm$ 10.5 & 63.0 $\pm$ 5.2  & 59.3 $\pm$ 5.2 \\
\bottomrule
\end{tabular}
\end{table}


\section{Conclusions}

In this paper, we study whether an LLM-driven agent can autonomously adapt honeypot exposure under strict resource constraints by inferring attacker intent from partial security telemetry. We formulate adaptive honeypot deployment as a sequential inference problem and evaluate the approach using a discrete-time simulator modeling multi-stage attacks aligned with the MITRE ATT\&CK framework.

Our results show that LLM-based agents can infer attacker progression and select which services to expose to sustain engagement. Performance degrades as attack-surface ambiguity increases, but stronger language models achieve higher attack-stage inference accuracy, indicating that semantic reasoning over IDS alerts can capture attacker objectives despite partial observability.
We further observe that exploitation success alone can be misleading: attackers may progress even when the agent’s internal estimate of the attack state is inaccurate. This highlights the need to jointly evaluate engagement and inference accuracy when assessing adaptive deception systems.

This work is preliminary and relies on scripted attackers and simplified abstractions of concurrent activity. Ongoing work focuses on validating the architecture in a live honeynet, extending the inference model to handle multiple concurrent attack hypotheses, and assessing whether fine-tuned local LLM models can achieve performance comparable to larger models. Overall, these results provide initial evidence that agentic AI can act as a viable control plane for adaptive cyber deception under realistic constraints.


\bibliographystyle{IEEEtran}
\bibliography{main}

\section*{Appendix}\label{sec:appendix}
\subsection{Deployed Services and Attack Phases}

This appendix describes the vulnerable services exposed in our simulated honeynet and
summarizes the corresponding exploitation chains implemented by the attacking proof-of-concept
(POC) exploits.

\subsubsection{Docker}
An exposed Docker daemon API can enable remote code execution (RCE) when unauthenticated access
to the API is permitted. In this case, an attacker may execute arbitrary container commands,
spawn privileged containers, and obtain root-level access on the host.
Since Docker containers often run with elevated privileges or direct access to host resources,
successful exploitation of the Docker API typically results in full host compromise, making
this attack vector particularly severe in cloud and containerized environments.

\subsubsection{Apache Struts}
CVE-2018-11776 is a remote code execution vulnerability affecting Apache Struts.
The attack sequence begins with identifying a Struts-based web application, followed by
directory traversal and endpoint enumeration to locate vulnerable paths, and culminates in
the execution of the exploit payload.
This exploitation chain illustrates the risks associated with legacy enterprise web frameworks,
where a single vulnerable endpoint can lead to full system compromise due to insufficient input
validation and insecure default configurations.

\subsubsection{GitLab}
CVE-2021-22205 affects GitLab CE/EE versions starting from 11.9 and enables unauthenticated
remote code execution due to improper validation of image files passed to a file parser.
Once initial access is obtained, an attacker may exfiltrate sensitive data and escalate
privileges, potentially resulting in full compromise of the GitLab instance.
Given GitLab’s central role in managing source code repositories and CI/CD pipelines, this
attack highlights the amplified impact of compromise in DevOps environments.

\subsubsection{Xdebug}
This attack models a common exploitation sequence against the PHP debugging extension Xdebug
when remote debugging is misconfigured and exposed in production environments.
An unauthenticated HTTP request triggers the DBGp debugging session, causing the server to
connect back to the attacker, who can then issue debug commands to execute arbitrary PHP code.
From this foothold, the attacker may perform data exfiltration, escalate privileges (depending
on web-server permissions), and ultimately compromise the host system.
Because Xdebug is intended for development use, its presence in production systems often
indicates poor deployment hygiene and significantly increases attack surface exposure.

Table~\ref{tab:details_exploitations} summarizes the exploitation phases implemented for each
deployed service in our simulation.

\begin{table}[ht]
\centering
\small
\begin{tabular}{lccccc}
\toprule
\textbf{Service} & \makecell{\textbf{Scan}} &
\makecell{\textbf{Initial}\\\textbf{Access}} &
\makecell{\textbf{Data}\\\textbf{Exfil}} &
\makecell{\textbf{Priv.}\\\textbf{Esc.}} &
\makecell{\textbf{Root}\\\textbf{Exfil}} \\
\midrule
GitLab        & \greencheck & \greencheck & \greencheck & \greencheck & \greencheck \\
Xdebug        & \greencheck & \greencheck & \greencheck & \greencheck & \greencheck \\
Apache Struts & \greencheck & \greencheck & \reddash    & \greencheck & \greencheck \\
Docker API    & \greencheck & \greencheck & \greencheck & \reddash    & \reddash \\
Others        & \greencheck & \reddash    & \reddash    & \reddash    & \reddash \\
\bottomrule
\end{tabular}
\caption{\textbf{Implemented attack phases per deployed service.} Green checkmarks indicate supported phases, while red dashes denote phases that are not modeled}
\label{tab:details_exploitations}
\end{table}

\end{document}